\documentclass[twocolumn,aps,preprintnumbers,amsmath,amssymb,floatfix,pre,showpacs]{revtex4}

\usepackage{dcolumn}
\usepackage{bm}
\usepackage{graphicx}

\providecommand{\remark}{%
\begin{rem}
    \normalfont}\def
    \endremark{%
\end{rem}
}

\providecommand {\R} {\ensuremath{\mathbb{R}}}

\providecommand{\pro}[1]{(#1_t)_{t\geq0}}

\providecommand{\ex}{\mathbb{E}}

\newcounter {constant}
\newenvironment{constant}{\refstepcounter{constant} }{}

\begin{document}

\makeatletter
\renewcommand\@dotsep{10000}
\makeatother

\pacs{02.50.Ey, 02.50.Ga, 05.40.Fb, 05.70.Fh}

\title
{\large Transition in the decay rates of stationary distributions of L\'evy motion in an energy landscape}
\vspace{1.5cm}
\author{
\small Kamil Kaleta \\[0.1cm]
 {\small\it    Institute of Mathematics, University of Warsaw}    \\[-0.4ex]
  {\small\it Banacha 2, 02-097 Warsaw, Poland}      \\[-0.4ex]
 {\small  {\tt kamil.kaleta@pwr.edu.pl}   }\\[0.5cm]
\small J\'ozsef L\H{o}rinczi\\[0.1cm]
{\it \small Department of Mathematical Sciences, Loughborough University} \\[-0.4ex]
{\it \small Loughborough LE11 3TU, United Kingdom} \\[-0.4ex]
{\small {\tt  J.Lorinczi@lboro.ac.uk}} \\[-0.4ex]}

\begin{abstract}
\noindent
The time evolution of random variables with L\'evy statistics has the ability to develop jumps, displaying
very different behaviors from continuously fluctuating cases. Such patterns appear in an ever broadening
range of examples including random lasers, non-Gaussian kinetics or foraging strategies. The penalizing
or reinforcing effect of the environment, however, has been little explored so far. We report a new phenomenon
which manifests as a qualitative transition in the spatial decay behavior of the stationary measure of a jump
process under an external potential, occurring on a combined change in the characteristics of the process and
the lowest eigenvalue resulting from the effect of the potential. This also provides insight into the fundamental
question of what is the mechanism of the spatial decay of a ground state.
\end{abstract}

\maketitle

\section{L\'evy statistics in physics}

Jump processes and related objects (L\'evy flights, continuous time random walks, heavy-tailed probability distributions,
non-local operators, fractional differential equations) are increasingly studied in mathematics and the natural sciences
due to their interest in modelling a multitude of phenomena, ranging from physics, through chemistry and biology, to social
science. Surveys with a wide scope include \cite{bib:BG,bib:SZU95,SKZ99,bib:MK1,bib:BDST,bib:ZDK}.

For a long time, Brownian motion played a central role in these modelling efforts, which has complex but continuous
sample paths, and other appealing statistical features such as moments of any order and a scaling property. These
ideas are now significantly extended by jump L\'evy and L\'evy-type (e.g., Feller) processes, whose application allows to take
account of further features such as sudden changes, spiky time series, heavy-tailed distributions, and are increasingly
recognized to provide life-like additions and refinements to the previously used concepts and tools, leading to a
new paradigm in scientific modelling. Due to the complexity of the problems, and the difficulties and subtleties
accompanying them (possible non-existence of statistical parameters such as any moments, non-locality, long range
memory effects) there is a strong need to have an ongoing communication between mathematics and physics. To address
these challenges, a concurrent mathematical research on the theoretical foundations of these new techniques reached a
high level of development \cite{bib:Sat,bib:BMR,bib:J,bib:AD,bib:MS,bib:Kol,bib:BSW}.

Recall that by a L\'evy process one understands a stochastic process $\pro X$ with independent increments in the
sense that if $0 < r < s < t$ are any distinct time points, then the random displacement $X_t-X_s$ is stochastically
independent from $X_s - X_r$. (Here $X_t$ is to be understood as a position in physical or a value in phase space,
taken at time $t$ at random with some distribution.) Furthermore, the increments are also stationary in the sense that
the probability distribution of the random displacement $X_t- X_s$ only depends on the difference $t-s$,
which means that displacements over equally long time intervals are identically distributed. Thirdly, with probability
one the paths $t \mapsto X_t$ of a L\'evy process are continuous from the right with left limits, i.e., $X_{t+} = X_t$
but it is not necessarily the case that $X_{t-} = X_t$, so a jump at $t$ may occur. (Here $t-$ means the moment of
time infinitesimally preceding $t$, and $t+$ means the moment infinitesimally soon after $t$). In fact, as it turns out,
L\'evy processes having jump discontinuities are the rule rather than the exception, and Brownian motion is rather the exceptional
L\'evy process which has continuous paths with probability one. Due to this special role of Brownian motion, the paths
of a L\'evy process $\pro X$ can in general be decomposed into four independent components so that $X_t = X_t^{(1)} +
X_t^{(2)} + X_t^{(3)} + X_t^{(4)}$, where $X_t^{(1)}$ is the drift component of the form $bt$ (with drift coefficient
$b$) having the effect of building in a non-random tendency into the full process, $X_t^{(2)}$ is the Brownian component,
$X_t^{(3)}$ is the small-jump component (with jump size up to 1), and $X_t^{(4)}$ is the large-jump component (with
jump size larger than 1). This decomposition implies that the paths of a general L\'evy process contain stretches of
continuous but very rugged curves, interspersed by jumps of any size and direction occurring at random times.

The frequency of jumps is described by the so-called L\'evy measure $\nu(z)dz$, constructed in the following way. A jump
at time $t$ occurs when $\Delta X_t = X_t - X_{t-}$ is non-zero, thus the total number of jumps up to time $t$ falling
in a $d$-dimensional box $E$ can be counted by $N(t,E) = \# \{s \in (0,t]: \Delta X_s \in E\}.$ By general theory,
it is then known \cite{bib:BMR} that the random number $N(t,E)$ follows a Poisson distribution whose mean is $t\nu(E) =
t\int_E \nu(z)dz$, thus $\nu(z)$, $z \in E$, measures the frequency of jumps of size in $E$. Moreover, the L\'evy measure satisfies
the integrability condition $\int_{\R^d\setminus \{0\}} \min\{1,|z|^2\} \nu(z) dz < \infty$, which means that the total activity
$\int_{|z|  > 1}\nu(z)dz$ of the large jumps is finite. 
Note that this does not rule out the possibility that the small-jump activity $\int_{0 <|z| \leq 1}\nu(z)dz$ is infinite.

All of the independent components of the full process contribute by a separate term into the exponent $\Psi$ of the Fourier
transform (characteristic function) of $X_t$, i.e., $\ex^0 \left[e^{i y \cdot X_t}\right] = e^{-t \Psi(y)}$, which is given by
the well-known L\'evy-Khintchine formula \cite{bib:BMR}:
\begin{eqnarray}
\label{LK}
\Psi(y)
&=&
-i b \cdot y + \frac{1}{2} y \cdot Ay \\
&& \quad + \int_{\R^d\setminus \{0\}}(1 - e^{i z \cdot y} + i y \cdot z 1_{\{|z| \leq 1\}}(z)) \nu(z)dz. \nonumber
\end{eqnarray}
Here $\ex^x[...] = \int ... \, d{\Bbb P}^x$ denotes expectation (average) with respect to the paths of the process
starting in $x \in \R^d$, distributed by the path measure ${\Bbb P}^x$, and $1_E(z)$ is the indicator function of set $E$,
i.e., it equals 1 if $z \in E$ and 0 if $z \not\in E$. The vector $b$ is the drift coefficient, the matrix $A$ is the diffusion
matrix, and $\nu(z)dz$ is the L\'evy measure.

L\'evy processes are widely encountered in physics. 
A first observation of the utility of Brownian motion derives from Feynman's path integration approach to quantum mechanics.
Making a Wick rotation $t \mapsto it$, the Schr\"odinger equation $i\partial_t \psi(x,t) = (-L + V)\psi(x,t)$, where $L =
\frac{1}{2}\Delta$ is half the Laplacian on $L^2(\R^d)$ and a system of units is adopted in which $\hbar = 1$, is transformed
into a diffusion equation with a dissipation function given by $V$. Then, with initial wave-function $\psi(x,0) = \phi(x)$, the
solution can be represented as (see \cite{bib:S,bib:Roe,bib:LHB} for details)
\begin{eqnarray}
\label{FKf}
\psi(x,t)
&=&
(e^{-t(-L+V)} \phi)(x) = \int_{\R^d} T(x,0 ; y,t) \phi(y) dy \nonumber \\
&=&
\ex^x\left[e^{-\int_0^t V(B_s)ds} \phi(B_t)\right],
\end{eqnarray}
where $\ex^x$ denotes averaging with respect to the paths of Brownian motion $\pro B$. The right hand side thus gives a stochastic
representation of the time-evolution semigroup $T_t =  e^{-t(-L+V)}$ with propagator $T(x,0;y,t)$, and the problem of studying the
solutions of the Schr\"odinger equation is now a problem of statistical mechanics in which a random mover is performing Brownian
motion in the potential landscape $V$.

A similar Feynman-Kac type representation holds for the semi-relativistic Hamiltonian with kinetic term $-L = (-\Delta + m^2)^{1/2}
- m$, involving the fractional Laplacian of order $\frac{1}{2}$ and where $m \geq 0$ is the rest mass of the particle \cite{bib:Ca,
bib:CMS,bib:TCB}. Then the corresponding semi-relativistic Schr\"odinger equation transforms into an anomalous (i.e., non-Gaussian)
diffusion equation, and Brownian motion is replaced by a jump L\'evy process, specifically, a Cauchy process with or without mass. When
the mass is non-zero, the L\'evy measure behaves like $\nu(z) \sim e^{-m|z|}|z|^{-d-1}(1+|z|)^{d/2}$, i.e., the frequency of jumps
decreases in leading order exponentially with their size. When the mass is zero, $\nu(z) \sim |z|^{-d-1}$, i.e., the jump size
distribution has polynomially heavy tails. (Here and below $\sim$ denotes large $|z|$ asymptotic comparability).

Quantities obeying L\'evy statistics occur also in many other models or experiments in physics.
The Cauchy process is the time evolution of stable random variables with index $\alpha = 1$, while other values of $\alpha$ have
further relevance. The Breit-Wigner formula for high-energy resonances is another example of occurrence of a symmetric Cauchy
distribution \cite{bib:BW}. Particles travelling between a point source and a detection plane produce an asymmetric Cauchy
distribution, uniformly distributed stars in space give rise to a gravitational field with stable distribution of index $\alpha = 1.5$,
the size of large polymerized molecules a distribution with the same index, the flight time of particles trapped in the vortices of a
flow is described by $\alpha = 1.3$, and so on \cite{bib:W}. In laser cooling technologies, based on making atoms accrue and
get trapped in low-momentum regions of the phase space, the trapping time is a stable random variable with $\alpha = 1.5$, and the
recycle time (the time needed to return to the trapping region), is a quantity with $\alpha = 1.25$ \cite{bib:BBAC,bib:BB}, see also
\cite{bib:KB}. Laser optics also provides L\'evy
statistical behaviors involving exponentially light-tailed jump measures. In \cite{bib:BBW} it has been reported that a medium of L\'evy
scatterers has been experimentally manufactured to produce a ``L\'evy glass", i.e., an optical material in which light travels
by following the rules of a L\'evy flight. In order to implement finite size requirements, so-called truncated L\'evy processes
have been proposed \cite{bib:MaS,bib:K}, which set hard cutoffs on jump sizes. A variant of these models works with soft cutoffs
by exponentially suppressing large jumps, involving L\'evy measures of the form $|z|^{-\delta}e^{-\beta|x|}$ \cite{bib:CdCN,bib:RJ}.
In a recent development it has been found that, since excitation pulses carry finite energies, in random lasers the output
intensities can be statistically described in terms of such truncated and exponentially tempered L\'evy variables, where parameter
fittings yielded the values $\delta = 0.53$, $\beta = 0.0054$ for unscattered light, and other truncations for the scattered light
component \cite{bib:UM}. Exponentially truncated situations have been observed also in sub-diffusion models \cite{bib:MZB}. Purely
exponential and Weibull distributions are supplied from extreme value phenomena, used in the description and prediction of the
behavior of e.g. maxima of a time-series, related to natural catastrophes, financial crashes, technological accidents etc
\cite{bib:EKM}.
The examples are further multiplied to a great extent by models of anomalous transport theory
\cite{bib:BG,bib:MK1,bib:MK3,bib:KRS,bib:CTB}, statistical physics \cite{bib:CKG,bib:CGK,bib:BSo,bib:GS1,bib:GM}, chemistry
\cite{bib:ZK,bib:MK2}, biology \cite{bib:BLVC,bib:SH,bib:PCM}, or social systems \cite{BHG,GS}.

\section{L\'evy motion in an energy landscape}

In order to unify the wide range of natural phenomena leading to L\'evy statistics in a single framework, we now consider a
model of a particle performing a spherically symmetric $\R^d$-valued L\'evy motion $\pro X$, with diffusion matrix $A = a I_d$
(where $I_d$ is the identity matrix in $\R^d$) and L\'evy intensity of the generic form
\begin{equation}
\label{generic}
\nu(z) = \nu(|z|) \sim e^{-c |z|^{\beta}} |z|^{-\delta},
\end{equation}
with parameters $c >0$, $\beta, \delta \geq 0$, under the effect of an external potential $V$. Our goal in this paper is to
study the stationary behavior of such a random mover in function of the jump measure and the potential. Specifically, we
focus on the spatial decay properties of the stationary measures, and report a qualitative transition in the decay rates as
the jump measure is varied from having a heavy to a light tail. Although in our argument we motivated the statistical mechanical
problem by the Feynman-Kac formula-based approach to non-relativistic and relativistic quantum mechanics, it should be clear
from the above that there are many anomalous kinetic models using non-local operators and L\'evy processes to which our framework
can be applied. Currently there are many efforts trying to find solutions of non-local equations, which is a difficult problem.
Precise information on the asymptotic behavior of eigenfunctions of non-local operators is important in verifying conjectures
and the validity of numerical approximations.

We distinguish the following three categories of jump measures:
\begin{enumerate}
\item
\emph{sub-exponential L\'evy measures:} $\beta \in [0,1)$, where $\beta = 0$ corresponds to purely polynomial
decay, and $\beta \in (0,1)$ corresponds to stretched exponential decay in leading order
\item
\emph{exponential L\'evy measures:} $\beta = 1$
\item
\emph{super-exponential L\'evy measures:} $\beta > 1$.
\end{enumerate}
These jump types cover a large number of L\'evy processes of interest. Apart from the examples discussed above, there
are many other processes which fit in one of these classes (jump-diffusions, geometric L\'evy processes, a variety of
other subordinate Brownian motions, Lamperti-type processes, hyperbolic processes etc). Due to a widespread occurrence
of scale-invariant phenomena and self-similar structures (fractals, ubiquitous power law-distributed data from natural
and social empirical observations etc), and of the theory of regular variation in mathematics, there is a rich
understanding of the sub-exponential class to date, while exponential and super-exponential cases were less explored
so far \cite{bib:AS,bib:AB}. An interesting comparative study of these three types has been made in \cite{bib:IPW,bib:BS},
analyzing the transition times in the Arrhenius-Eyring-Kramers laws of chemical reaction rates, revealing the differing
escape mechanisms through a potential barrier in the weak noise limit, according to the specific choices of the driving
process. We discuss here differing qualitative behaviors occurring around a sharp transition point of similar jump
processes, at arbitrary noise strengths.

With these entries the L\'evy-Khintchine formula (\ref{LK}) reduces to
\begin{equation}
\label{LKhere}
\Psi(y) = \frac{a}{2}|y|^2 + \int_{\R^d \setminus \{0\}} (1-\cos(y \cdot z)) \nu(z)dz,
\end{equation}
and in the path decomposition described above we have a Brownian component with diffusion coefficient $a \geq 0$, and
(small and large) jump components with L\'evy measure $\nu(z)dz$, while the assumed symmetry prevents a drift term to occur.
Spherical symmetry implies $\nu(z)= \nu(|z|)$ and thus the process can make a jump of size $|z|$ in any direction with equal
likelihood.

Next we consider such a random mover in an energy landscape described by a potential $V$, acting as a mechanism reinforcing
or penalizing the random mover to go in specific regions of space.
We assume that $V = V^+ - V^-$ is Kato-regular in the sense that its negative part satisfies
$$
\lim_{t \rightarrow 0} \sup_{x \in \R^d} \ex^x \left[\int_0^t |V^-(X_s)| ds\right] = 0,
$$
and the restriction to every ball in $\R^d$ of its positive part $V^+$ satisfies the same condition. This condition implies
by the Markov property of the process that $-\int_0^t V(X_s) ds$ is exponentially integrable for all $t \geq 0$, and thus the
Feynman-Kac formula holds. On the other hand, it is general enough to accommodate potentials which may have local singularities
and any number of local minima or maxima. The crucial fact is that the long time behavior of the process depends on the
asymptotic behavior of the potential at infinity. We consider the following two large classes:
\begin{enumerate}
\item
\emph{confining potentials:} with $\lim_{|x|\to\infty}V(x) = \infty$
\item
\emph{decaying potentials:} with $\lim_{|x|\to\infty}V(x) = 0$.
\end{enumerate}
Examples of confining potentials include the quadratic (harmonic) and all the even anharmonic potentials, and some decaying
potentials are potential wells and Yukawa potentials (there is a large supply of further examples in either class).

In the model we consider, the potential conditions the L\'evy motion $\pro X$ turning it into a new process $\pro {\widetilde X}$.
Under $V$ correlations are introduced between different values $\widetilde X_s$, $\widetilde X_t$, hence this is not a L\'evy
process. Notwithstanding this crucial difference, it is still a Markov process. By multiplication through the Feynman-Kac factor
$e^{-\int_0^t V(X_s)ds}$, the path measure in the expectation at the right hand side of (\ref{FKf}) loses its normalization, so $T_t$
is not a stochastic semigroup. To determine the stochastic semigroup describing $\pro {\widetilde X}$, consider the ground state
$\varphi_0$ of the Hamiltonian $-L+V$, i.e., the unique eigenfunction lying at the lowest eigenvalue $\lambda_0$ of the Hamiltonian,
whenever this exists. Note that the ground state has no nodes and $\varphi_0(x) > 0$ for all $x \in \R^d$. A combination of the
eigenvalue equation and (\ref{FKf}), and division at both sides by $\varphi_0(x)$ gives
$$
1 = e^{\lambda_0t}  \int_{\R^d} T(x,0 ; y,t) \frac{\varphi_0(y)}{\varphi_0(x)}dy =  \int_{\R^d} \widetilde T(x,0 ; y,t) dP(y),
$$
with
$$
\widetilde T(x,0 ; y,t) = \frac{e^{\lambda_0t} T(x,0 ; y,t)}{\varphi_0(x)\varphi_0(y)}
$$
and $P$ given by eq. (\ref{sta}) below. Thus the semigroup $\widetilde T_t$ with propagator $\widetilde T(x,0 ; y,t) > 0$ for all
$x,y \in \R^d$ and $t > 0$, satisfies the normalization $\widetilde T_t 1 = 1$ at all time, and so it is a stochastic semigroup
with stationary measure
\begin{equation}
dP(x) = \varphi_0^2(x)dx.
\label{sta}
\end{equation}
A further computation \cite{bib:KL12,bib:GS2} gives the transition probability density of the random process $\pro {\widetilde X}$
from position $x$ at time $s$ to a position $y$ at time $t$ to be
\begin{eqnarray}
\label{FK}
&& \widetilde p(x,s \, | \, y,t) \\
&& \ \ \ =  \frac{e^{\lambda_0(t-s)}}{\varphi_0(x)\varphi_0(y)} \,
\ex^{x,y}_{s,t}[ e^{-\int_s^t V(\widetilde X_r)dr}] \, p(x,s \, | \, y,t), \nonumber
\end{eqnarray}
where $p(x,s \, | \, y,t)$ is the transition probability density of the underlying process $\pro X$, and the notation indicates
conditional expectation with pin-down conditions $\widetilde X_s = x$ and $\widetilde X_t = y$. The path measure of the process
under $V$ can also be calculated, and we obtain that the probability that $\pro {\widetilde X}$ starting from $x$ will be in a
region $A$ in space at time $t$ is
\begin{equation}
\label{gibbs}
\widetilde {\mathbb P}^x(\widetilde X_t \in A) = \int_A \widetilde p(x,0 \, | \, y,t) \varphi_0^2(y)dy.
\end{equation}
In light of \eqref{FK}, the right hand side shows that the path measure can be interpreted as a Gibbs distribution over the paths
of the random process, for the energy function given by the integral of $V$.

The infinitesimal generator of $\pro {\widetilde X}$ is (see also \cite{bib:PZ})
\begin{eqnarray*}
(\widetilde L f)(x) && = \frac{a}{2} \Delta f (x) + a \nabla\log\varphi_0(x) \cdot \nabla f(x) \\ && + \int_{0<|z| \leq 1}
\frac{\varphi_0(x+z)-\varphi_0(x)}{\varphi_0(x)} \, z \cdot \nabla f(x) \, \nu(z)dz   \\
&& + \int_{\R^d\setminus \{0\}}\big(f(x+z) - f(x) \\
&& \ \ \ \ \  - z \cdot \nabla f(x) 1_{\{|z| \leq 1\}}(z)\big) \frac{\varphi_0(x+z)}{\varphi_0(x)}\nu(z)dz. \nonumber
\end{eqnarray*}
The expression indicates that the jump measure and drift are now position dependent (unlike for a L\'evy process), and the process
$\pro {\widetilde X}$ will have different jump preferences according to what the value of $V$ at $x$ is.
Note that the effect of the potential comes in via the ground state $\varphi_0$. Interestingly, for the process constrained by the
potential two types of drift appear. The first takes the usual form and is related to the diffusive part of the underlying process.
The second corresponds to the jump part of the underlying L\'evy process and is given by the integral difference operator. While
now the motion is in some sense more complex and loses its isotropy, the topology of the paths is preserved on applying the potential,
in the sense that Brownian motion under $V$ keeps having continuous paths, and a jump L\'evy process under $V$ keeps having continuous
stretches interrupted by random jumps.

\section{Stationary behavior}

In general, when $V \equiv 0$ the process $\pro X$ will display wild fluctuations, exploring the full space. Brownian motion has no
$t \to \infty$ limiting behavior in the sense that $\liminf_{t\to\infty} B_t = -\infty$ and $\limsup_{t\to\infty} B_t = \infty$
with probability one. However, the law of iterated logarithm says that with probability one $|B_t| \leq \sqrt{2t\log\log t}$,
after a sufficiently long time. In other words, Brownian paths eventually get localized with a high probability inside a curved
cone given by the iterated log-profile function. For jump L\'evy processes which have infinite variance (such as stable
processes), the situation is very different as there exists no such profile function and no region in space in which paths concentrate
on the long run.

However, when a non-zero potential is applied to a L\'evy process, the long term behavior dramatically changes. Paths will now tend
to spend long times at the local/global minima of the energy landscape since (\ref{FK}) implies that strays involve an
exponentially high price. On the long run, the process will tend to settle on a behavior described by the stationary measure $P$
given by (\ref{sta}), whenever this exists, thus the key to the stationary behavior of the process $\pro {\widetilde X}$ is the ground
state $\varphi_0$.
Explicit calculations of ground states for
L\'evy processes are known so far for $V(x) = x^2$ \cite{LM12} and $V(x) = x^4$ \cite{DL15}.
The tail behavior of $P$, which amounts to the asymptotic behavior of $\varphi_0$ as $x \to \pm \infty$, tells
then of how strongly the random mover is localized in the bulk, in particular, around the local and/or global minima of the potential.

A ground state, and thus a stationary behavior of the process conditioned by $V$, occurs only if the potential is suitably chosen.
When the potential is confining, a ground state always exists. When the potential is decaying, it is in general a hard problem
to determine whether for a given L\'evy process and a given potential a ground state exists, and it is possible to construct
many examples of jump processes and potentials for which it does not \cite{bib:CMS,bib:LL}. The physical reason why a ground state
$\varphi_0$ forms at all is due to a mechanism allowing the process to accumulate a sufficient mean total sojourn time in each
unit-ball neighborhood of every point in space through visiting and revisiting them \cite{bib:DV,bib:R}. Clearly, if $V$ is confining,
then (\ref{FK}) implies that far out the paths will be exponentially penalized and a strong centripetal effect drives most of the
motion to take place near the bottom region of the energy landscape, which makes the process steady down to a stationary state.
If $V$ is decaying, then far out the motion increasingly resembles free motion (which has no ground state) and the energetic effect
becomes more delicate so that a ground state forms for a sufficiently low-lying ground state eigenvalue only.

Whenever a ground state does exist, we can derive a useful representation by using the eigenvalue equation. Since
the ground state eigenvalue $\lambda_0$ may be negative dependent on the choice of potential, we shift it by
$\theta$ so that $\theta + \lambda_0  >0$. Then on multiplying both sides of the eigenvalue equation by $e^{-\theta t}$
and integrating with respect to time, we get
\begin{align}\label{eq:eig}
\varphi_0(x) = (\theta+\lambda_0)\int_0^{\infty}\ex^x\left[ e^{-\int_0^t(\theta+V(X_s))ds} \varphi_0(X_t) \right]dt.
\end{align}
We can manipulate the integral at the right hand side by choosing an arbitrary bounded open set $D$ in space, containing
the starting point of the process, and considering the first exit time
$$
\tau_D = \inf\{t \geq 0: \, X_t \not\in D\}
$$
of the process from $D$. Using this, the process can be stopped at this random time and then allowed to run from this time
onwards by using its Markov property. By splitting up the time integral accordingly, this leads to
\begin{eqnarray*}
\label{gssplit}
\lefteqn{
\varphi_0(x) = (\theta+\lambda_0) \times } \\
&& \times \, \ex^x\left[\left( \int_0^{\tau_D} + \int_{\tau_D}^{\infty} \right) e^{-\int_0^t(\theta+V(X_s))ds}
\varphi_0(X_t) dt \right] \\
&=&
(\theta+\lambda_0) \ex^x\left[ \int_0^{\tau_D} e^{-\int_0^t(\theta+V(X_s))ds} \varphi_0(X_t) dt\right] \\
&&
\quad + \, \ex^x\left[1_{\{\tau_D < \infty\}} e^{-\int_0^{\tau_D}(\theta+V(X_s))ds} \varphi_0(X_{\tau_D}) \right].
\end{eqnarray*}
It is in our gift how to choose $D$ optimally to study the large $|x|$ behavior of the ground state, and we take it
to be a unit ball $B_1(x)$ centered in $x$. Making use of the above representation, after some analysis we find that
the second term provides the main contribution into a lower bound of $\varphi_0(x)$, and the first term in an upper
bound. Putting these estimates together, we obtain
\begin{equation}
\label{meantime}
\varphi_0(x) \sim
\ex^x\left[\int_0^{\tau_{B_1(x)}} e^{-\int_0^t V(X_s)ds} \right] \, \nu(x)
\end{equation}
for a large class of L\'evy processes $\pro X$, where $\tau_{B_1(x)}$ is the first exit time from this ball by the process
starting at $x$. 
The factor multiplying $\nu(x)$ in (\ref{meantime}) measures the mean time spent by $\pro  X$ perturbed by the potential $V$ in a
radius one neighborhood of $x$ before the first exit from this region. This shows that while the creation of a ground
state depends on the right amount of time being spent in specific locations of space, the decay of a ground state is governed
by another mechanism, which determines how soon the random mover leaves such unit-balls far out \cite{bib:KL15a,bib:KL15b}.


If $V$ is a confining potential with a mild additional property, we find through an analysis of the mean exit times in
(\ref{meantime}) that whenever $\nu$ is sub-exponential or exponential with $\delta  > \frac{d+1}{2}$, the ground state
behaves like
$$
\varphi_0(x)  \sim \frac{\nu(x)}{V(x)}.
$$
This gives a neat account of the separate contributions of the potential and the underlying free process, and
the expression shows that the decay rate of $\varphi_0$ results from a balance between the L\'evy intensity
and the killing effect of the potential far out.

If $V$ is a decaying potential, the situation becomes more subtle since now both $\nu$ and $V$ go to zero and a
similar balance does not hold. The effect of the potential comes in now through the negative ground state
eigenvalue $\lambda_0$ it produces (if any). The following behaviors occur:
\begin{enumerate}
\item[(a)]
\emph{polynomially decaying L\'evy intensities:} if $\nu(|z|) \sim |z|^{-\delta}$ with $\delta >d$,
then $\varphi_0(x) \sim |x|^{-\delta}$ for all $\lambda_0 < 0$

\item[(b)]
\emph{stretched-exponentially decaying L\'evy intensities:} if $\nu(|z|) \sim e^{-c |z|^{\beta}}
|z|^{-\delta}$, with $c >0$, $\beta \in (0,1)$, $\delta \geq 0$, then $\varphi_0(x) \sim
e^{-c|x|^{\beta}} |x|^{-\delta}$ for all $\lambda_0 < 0$

\item[(c)]
\emph{exponentially decaying L\'evy intensities:} if $\nu(|z|) \sim e^{-c |z|} |z|^{-\delta}$ with
$c >0$ and $\delta > \frac{d+1}{2}$, then there is a process-dependent cutoff $\eta_X$ such that whenever $|\lambda_0|
> \eta_X$, we have $\varphi_0(x) \sim  e^{-c|x|} |x|^{-\delta}$; when there is no restriction on the
eigenvalue $\lambda_0$, we have for large enough $|x|$ that for every $\varepsilon > 0$ there is a
$C_\varepsilon > 0$ such that
$$
\varphi_0(x) \geq C_\varepsilon \max \left\{e^{- \widetilde c \sqrt{|\lambda_0|+\varepsilon} |x|},
e^{-c |x|}|x|^{-\delta}\right\};
$$
moreover, if $\delta \leq \frac{d+1}{2}$, then $\varphi_0(x) > C\nu(x)$ for every constant $C > 0$, at
least for large enough $|x|$

\item[(d)]
\emph{super-exponentially decaying L\'evy intensities:} if $\nu(|z|) \sim e^{-c |z|^{\beta}}
|z|^{-\delta}$ with $c>0$, $\beta > 1$ and $\delta \geq 0$, then for every $\varepsilon > 0$ there is a
$C_\varepsilon > 0$ such that
$$
\varphi_0(x) \geq C_\varepsilon e^{- \widetilde c \sqrt{\min\{|\lambda_0|+\varepsilon, 1\}} \,
|x| (\log|x|)^{(\beta-1)/\beta}}
$$
for large enough $|x|$ and any $\lambda_0 < 0$

\item[(e)]
\emph{diffusions:} if $\pro X$ is Brownian motion, $V$ is 
bounded and continuous, then for all $0 < \varepsilon < |\lambda_0|$ there is
$C_\varepsilon  > 0$ such that \cite{bib:A}
$$
\varphi_0(x) \leq C_\varepsilon e^{-\sqrt{\frac{|\lambda_0|-\varepsilon}{2}} \,|x|}.
$$
\end{enumerate}

From the above behaviors it is seen that the ground state decay follows the decay of the L\'evy intensity for
sub-exponential cases and $\lambda_0$ which can be arbitrarily close to zero. For super-exponential cases the rate
of decay is slower than that of $\nu$, whatever the magnitude of $\lambda_0$, and this comparison can be further
continued to diffusions to find that $\varphi_0(x)$ decays at infinity much slower than $p(x,0 \,|\, 0,1)$, where
$p(x,s\,|\,y,t) = (2\pi (t-s))^{-\frac{d}{2}} e^{-\frac{|y-x|^2}{2(t-s)}}$ is the Gaussian transition probability
density which can be used instead of $\nu$.
The exponential case is an interesting and delicate borderline situation, marking a transition line in the decay
patterns. In this case also the magnitude of the ground state eigenvalue and the sub-leading order polynomial
factors in $\nu(x)$ play a role. Whenever $\delta > \frac{d+1}{2}$, the absolute value of $\lambda_0$ determines
what the decay rate of $\varphi_0$ is: if $|\lambda_0|$ is large enough, then the ground state follows the decay
of the L\'evy intensity, if $|\lambda_0|$ is too small, then $e^{- \widetilde c \sqrt{|\lambda_0|+\varepsilon} |x|}$
may become larger than $e^{-c |x|}|x|^{-\delta}$ and thus the decay of the ground state gets slower than that of the
L\'evy intensity. As soon as $\delta \leq \frac{d+1}{2}$, the decay becomes invariably slower than that of $\nu$, no
matter the value of $|\lambda_0|$.

These results are obtained through an analysis of specific preferred jump scenarios and the processes' response
to perturbation by the potential. A key observation is that the sub-exponential processes (with $\delta > d$ for
polynomial and $\delta \geq 0$ for stretched exponential $\nu$) satisfy the property that events in which two
arbitrary points are linked through a sequence of any length of successive large jumps via intermediary points,
are stochastically dominated by events consisting of a single direct large jump between the given two points.
This \emph{jump-paring property} has been explored in \cite{bib:KL15a, bib:KL15b, bib:KS15}. It fails to hold for
super-exponential processes, and there is a divide around the critical exponent $\delta = \frac{d+1}{2}$ for
exponential processes, above which it does hold, and at and below which it breaks down. When the jump-paring
property fails to hold, the ground state necessarily decays slower than the L\'evy intensity. The process-dependent
cutoff $\eta_X$ appearing in (c) above relates to how much energy needs to be fed to the process to make it ``bend"
and follow the decay of the intensity of jumps.

\section{Transition in the decay rates}

The regime change occurring at exponentially decaying L\'evy intensities can be explained by a switch-over in the
mechanism which makes the long jumps of the process occur. This can be seen by separating the large-jump component
$\pro {\overline {X}}$ from the full process $\pro X$. Its L\'evy intensity can be obtained by restriction of the
L\'evy intensity of the full process to large jumps, i.e., $\nu_\infty(z) = \nu(z) \rceil_{\left\{|z| \geq 1\right\}}$.
Using this, the jump-paring condition is formulated as
\begin{eqnarray*}
(\nu_\infty \ast \nu_\infty)(z-x)
&=&
\int_{\R^d} \nu_\infty(y-x)\nu_\infty(z-y) dy \\
&\leq&
C \nu_\infty(z-x),
\end{eqnarray*}
where the star indicates convolution and $C>0$ is a constant. This expresses the property that a large
jump from $x$ to $y$ and then on from $y$ to $z$ is less preferred than a direct large jump from $x$ to
$z$. Continued inductively, it follows that the $n$-fold convolution $\nu_\infty^{n\ast} \leq C^{n-1}
\nu_\infty$, so all multiple large jumps are stochastically dominated by single large jumps.

Since by the general integrability condition on the L\'evy intensity stated above $\nu$ is integrable
with respect to the large jumps, $\nu_\infty$ can be normalized to the probability measure
$\overline \nu_\infty(z)dz = \nu_\infty(z) dz/\int_{|z| \geq 1}\nu_\infty(z) dz$, which describes the
distribution of large jumps of size $|z|$. Denote the successive jumps of $\pro {\overline{X}}$ by
$J_1, J_2, J_3,...$ of sizes $|J_1|, |J_2|, |J_3|,...$  They are mutually independent random variables
described by the probability measure $\overline P(J_i \in dz) = \overline \nu_{\infty}(z) dz$. Whenever
$\pro X$ is a L\'evy process with sub-exponentially decaying $\nu$, we have for large enough $r$ that
$$
\overline P(|J_1 + J_2| > r) \approx 2 \overline P(|J_1| > r).
$$
A straightforward calculation using this property gives
\begin{eqnarray*}
\overline P(\max\{|J_1|, |J_2|\} > r) 
\approx \overline P(|J_1+J_2| > r)
\end{eqnarray*}
as well as
\begin{eqnarray*}
\overline P\left(\frac{r}{2} \leq |J_1| \leq r, \, \frac{r}{2} \leq |J_2| \leq r\right)
&=&  o(\overline P(|J_1| > r)).
\end{eqnarray*}
The implication is that a displacement in the random mover's position resulting from a double jump
$J_1,J_2$ is larger than a large value $r$ when either $|J_1| > r$ or $|J_2| > r$, and it is
much less likely that $r$ is exceeded cumulatively by jumps having comparable size. This
property extends to arbitrarily long sequences of jumps. When $\pro X$ has an exponentially
decaying $\nu$, this no longer holds and we have (with $\delta = 0$)
$$
\overline P\left(\frac{r}{2} \leq |J_1| \leq r, \, \frac{r}{2} \leq |J_2| \leq r\right)
\approx \overline P(|J_1| > r),
$$
i.e., the same order of magnitude, meaning that now a large displacement $r$ is achieved typically
cumulatively rather than through a single large jump. This can be seen as an increase of the capacity of
the process to fluctuate through multiple smaller jumps rather than exceedingly large single jumps, which
improves as the tail of $\overline \nu_\infty$ (and thus of $\nu$) is chosen lighter.

The conclusion one can draw from this transition phenomenon is that there are unexpected features in how
jump processes respond to even small perturbations by a potential. To appreciate this on a specific example,
consider a $d$-dimensional potential well $V(x) = -v 1_{\{|x| \leq a\}}$ of depth $v > 0$ and diameter $2a$,
and a random mover performing L\'evy motion $\pro X$ of the above jump categories, in the landscape of this
potential. In this case
$$
\int_0^t V(X_s) ds = -v \int_0^t 1_{\{|X_s| \leq a\}} ds = -v U_t(a),
$$
where $U_t(a)$ is the total time the process spends in the potential well within the time interval $[0,t]$. Since
in $d \geq 3$ every L\'evy process is transient and the mean total sojourn time $\ex^x [U_\infty(a)]$ is finite,
a ground state will form only for sufficiently large depths $v > v^*$ of the potential well, and the threshold value
$v^*$ depends on the process. When $d \leq 2$, then for some processes a ground state forms for arbitrarily small
non-zero $v$ (e.g., in the case of Brownian motion), while for others it may depend on further details (e.g., for a
spherically symmetric $\alpha$-stable process there is a ground state for every $v >0$ if $d=1$ and $1 \leq
\alpha < 2$, but only for large enough $v$ when $0 < \alpha < 1$, while for $d=2$ for every $\alpha$ a deep enough
potential well is needed). When a ground state does exist, the processes can be ranked according to their inherent
ability to form such a ground state, and we find that Brownian motion and processes with super-exponential L\'evy
measures form  a ground state easiest, and polynomially heavy-tailed processes the hardest \cite{bib:LL}.

Assuming that a ground state exists, consider now random motion according to a stable process, starting from the
origin, i.e., the middle of the potential well. Suppose that after some time the mover gets to a long distance $r$
from the origin. Due to the dominance of single large jumps, it is likely that it will get there by a single jump
of comparable size to $r$, without ``wasting" too much time with intermediary visits elsewhere. Before the next
large jump occurs, the process performs smaller fluctuations more locally. Since there is an energy gain in returning
to the region inside the potential well, this will happen with a good chance, however, just because another large
jump is due soon, the mover will not stay around zero for long. This back and forth motion repeats then at
a rate proportional to the jump intensity, and so the process develops a stationary distribution tracking
the curve of $\nu$, with their tail behaviors coinciding exactly. A similar pattern occurs for all motions
of sub-exponential and jump-paring exponential type (i.e., $\delta > \frac{d+1}{2}$).

Next consider processes with exponentially decaying jump intensity for which the jump-paring property no
longer holds ($\delta \leq \frac{d+1}{2}$). Now the paths from a long distance $r$ can cluster around the
origin less efficiently and once they are back, they spend comparatively large amounts of time in its
neighborhood, and re-entries at shorter gaps build up ``backlogs" in the decay-events so that (\ref{meantime})
does not hold any longer. Thus their ground states will decay possibly much slower than $\nu$. The same pattern
occurs for super-exponentially light jumps and continuously fluctuating processes (diffusions), which have strong
limitations to make long trips in short times. The accumulation of these backlog events makes a sudden qualitative
change to occur, which can be seen in the polynomial order sub-leading term at a critical exponent, which we can
identify explicitly to be $\frac{d+1}{2}$.

The property that the ground state follows the decay regime of the L\'evy intensity is a combined effect of
the jump-paring property favoring single large jumps over multiple large jumps, and the path concentrating
effect around the bottom of the well due to the ground state energy $\lambda_0$. The efficiency of this
mechanism depends on how low-lying $\lambda_0$ is. The above results show that when the process typically
evolves through large jumps, a small energetic effect suffices for securing an efficient concentration. As a
result, the hierarchy of good stationary behavior is then that the ground state of a jump process in a deep
enough potential well decays like $\nu$, decays slower than $\nu$ if the well is not deep enough, and the
ground state may even cease to exist if the well is too shallow.



\begin{acknowledgments}
\noindent
The authors thank IH\'ES, Bures-sur-Yvette, for visiting fellowships providing an ideal environment to joint
work. This research was partly supported by the National Science Centre (Poland) Grant on the basis of the
decision no. DEC-2012/04/S/ST1/00093 and by the Foundation for Polish Science.
\end{acknowledgments}

\end{document}